\documentclass[twocolumn]{aastex62}
\usepackage{hyperref,astrojournals,amsmath,amssymb,mathtools,graphicx,txfonts,microtype,nicefrac,xspace,bm}
\usepackage[switch]{lineno}
\usepackage{enumitem,kantlipsum}

\renewcommand*{\vec}[1]{\ensuremath{\boldsymbol{#1}}}
\let\oldhat\hat
\renewcommand*{\hat}[1]{\oldhat{\vec{#1}}}

\newcommand{\Md}{M_{\text{d}}}

\newcommand{\ts}{t_{\rm sec}}
\newcommand{\muvec}{\bm{\mu_{\hat{e}}}}
\newcommand{\mux}{\mu_{\hat{\bm{e}},x}}
\newcommand{\muy}{\mu_{\hat{\bm{e}},y}}
\newcommand{\muz}{\mu_{\hat{\bm{e}},z}}
\newcommand{\muR}{\mu_{\hat{\bm{e}},R}}

\renewcommand*{\vec}[1]{\boldsymbol{#1}}

\hypersetup{%
  pdftitle={Apsidal Alignment},
  pdfauthor={\textcopyright\ authors},
  bookmarksopen=true,
  colorlinks=true,
  linkcolor=black,
  citecolor=black,
  urlcolor=black}

\begin{document}

\title{Apsidal Clustering following the Inclination Instability}

\shorttitle{Apsidal clustering} 

\shortauthors{Zderic, Collier, Tiongco \& Madigan}
 
 \author{Alexander Zderic}
\affiliation{JILA and Department of Astrophysical and Planetary Sciences, CU Boulder, Boulder, CO 80309, USA}
\email{alexander.zderic@colorado.edu}

\author{Angela Collier}
\affiliation{JILA and Department of Astrophysical and Planetary Sciences, CU Boulder, Boulder, CO 80309, USA}

\author{Maria Tiongco}
\affiliation{JILA and Department of Astrophysical and Planetary Sciences, CU Boulder, Boulder, CO 80309, USA}

\author{Ann-Marie Madigan}
\affiliation{JILA and Department of Astrophysical and Planetary Sciences, CU Boulder, Boulder, CO 80309, USA}

\begin{abstract}
Disks of low-mass bodies on high-eccentricity orbits in near-Keplerian potentials can be dynamically unstable to buckling out of the plane. 
In this letter, we present $N$-body simulations of the long-term behavior of such a system, finding apsidal clustering of the orbits in the disk plane. 
The timescale over which the clustering is maintained increases with number of particles, suggesting that lopsided configurations are stable at large $N$.  
This discovery may explain the observed apsidal ($\varpi$) clustering of extreme trans-Neptunian Objects in the outer solar system.
\end{abstract}


\section{Introduction}
\label{sec:intro}

Collective gravity is responsible for most large-scale structure in disk galaxies, e.g. buckling, bars, and spiral arms.
This letter is a continuation of a series exploring analogous effects in the near-Keplerian potential of the Solar System. In \citet{Madigan2016} we presented a new dynamical instability driven by the collective gravity of low mass bodies in an axisymmetric near-Keplerian disk, and applied our results to the outer Solar System ($\sim100-1000$ AU).  This ``inclination instability'' exponentially grows the orbital inclinations of bodies while decreasing their orbital eccentricities, raising their perihelia and clustering their arguments of perihelion ($\omega$). It appears in many ways similar to the out-of-plane buckling instability of barred disk galaxies \citep{rahaetal,friedlipfenniger}. 
In \citet{Madigan2018b} we explained the mechanism behind the instability: long-term (secular) torques acting between high-eccentricity orbits, and showed it scaled with the number of particles in the simulation. In \citet{Fleisig2020}, we moved from simulations of a single mass population to a mass spectrum. 
In \citet{Zderic2020}, we showed that $\mathcal{O}(20\,M_{\oplus})$ is required for the instability to occur in a primordial scattered disk between $\sim100-1000$ AU in the solar system under the gravitational influence of the giant planets.
We also demonstrated how the instability naturally generates a gap in perihelion at a few hundred AU. 
The saturation timescale for the instability in a 20 Earth mass disk is $\lesssim 660$~Myr. Hence the non-linear, saturated state of the instability is important to understand. In this letter we look at the long-term behavior of the system and discover a new effect: apsidal clustering of orbits in the disk plane.

\section{Long-term evolution of the Inclination Instability}
\label{sec:results}

\begin{figure*}
    \centering
    \includegraphics[scale=0.35]{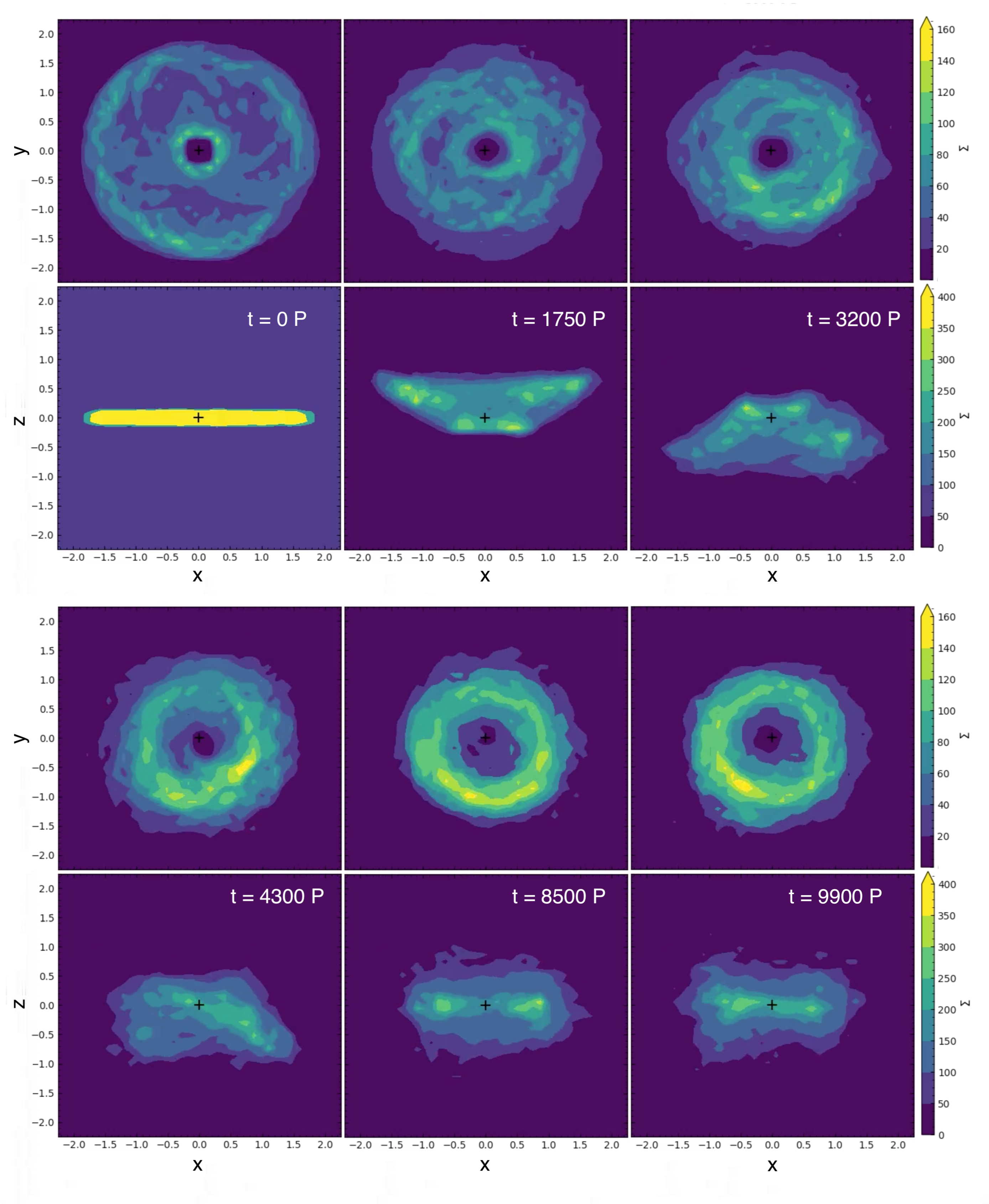}
    \caption{Surface density of the disk at different times showing mode development in $x$-$y$ plane (top rows) and $x$-$z$ plane (bottom rows). 
   	Orbital motion is CCW in the $x$-$y$ plane, and individual disk orbits precess CW.
    	Time progresses from top left to bottom right. 
    	The initially flat disk undergoes the inclination instability, buckling out above the $x$-$y$ plane and dropping in orbital eccentricity ($t \sim 1750\,P$).
	The orbits precess back through the plane, moving the `cone' of orbits below the $x$-$y$ plane ($t \sim 3200\,P$).
	The $m=1$ mode grows while the orbit cone disperses due to differential precession (from $t \sim 4300\, P$).
	This whole process takes $\sim 60\,\ts$.
    	The evolution of mean normalized eccentricity vector $\bm{\mu_{\hat{e}}}$ for this simulation is plotted as the stable model in Figure~\ref{fig:ecc-vec-evo}.}
    \label{fig:surface-density-plots}
\end{figure*}

We perform $N$-body simulations using the IAS15 adaptive time-step integrator in  \texttt{REBOUND} \citep{Rein2012}. 
We use \texttt{REBOUNDx} \citep{Tamayo2019} to add a zonal quadrupole ($J_2$) term to the potential of the central object to approximate the influences of the giant planets (Jupiter, Saturn, Uranus, and Neptune). 

In these simulations, we return to our simplified\added{, approximately mono-energetic} setup as in \citet{Madigan2016,Madigan2018b}. 
This \added{idealized} setup is chosen in place of a scattered disk configuration as higher particle density is needed to see the apsidal clustering (see section~\ref{sec:discussion}).
The disk of orbits is initialized with a semi-major axis $a$ distribution drawn uniformly in \replaced{$[0.9,1.1)$}{$[0.9,1.1]$}, eccentricity $e=0.7$, and inclination $i=10^{-4}\,{\rm rad}$, and is initially axisymmetric (argument of perihelion, $\omega$, longitude of ascending node, $\Omega$, and mean anomaly, $\mathcal{M}$, drawn from a uniform distribution in $[0,2\pi)$). 
\added{The large initial eccentricities are motivated by the solar system's scattered disk.}
\added{The Newtonian $N$-body problem is scale-free. This means we can apply our results to different semi-major axes by scaling the timescale. For example, if we choose $a = 1 = 100\,AU$, one orbital period corresponds to $P = 10^3$ years.}
The total mass of the disk is $\Md = 10^{-3}\,M$ and the number of disk particles, $N = 400$. 
The inclination instability scales with the secular timescale, 
\begin{align}
	\ts \sim \frac{M}{\Md} \frac{P}{2\pi}.
	\label{eq:sec-time}
\end{align}
\deleted{where $P$ is the orbital period at $a=1$.} 
With this definition, $\ts \approx 170\,P$.
\added{As in our previous publications, we deliberately simulate an unrealistically large mass ratio, $M/\Md$, to reduce the secular timescale and hence the simulation wall-time. To apply our results to the solar system, we rescale using the secular timescale. For example, the timescale for the instability to occur in a $\sim$20 Earth mass disk will be $\sim$16 times longer than in the simulations presented here.}

In Figure~\ref{fig:surface-density-plots} we show surface density evolution of the disk with face-on and edge-on lines of sight.  
The orbits incline out of the plane, collectively pitching over their semi-latus rectum and rolling over their major axis. Collectively, the orbits describe a cone shape. 
They drop in eccentricity as they incline, visibly contracting the surface area of the disk.
The orbits reach peak mean inclination at $\sim2000~P$, and we observe the formation of a prograde-precessing $m=1$ mode in the disk soon after.  
The mode starts in the inner disk as a single spiral arm (top right panel) and then moves to larger radii forming a banana-shaped over-density. 
 This is a slow mode \citep{Tremaine2001,Tremaine2005}, with a pattern speed of $\sim7 \times 10^{-4}\,$rad$\,$P$^{-1}$. 
In the bottom left panel we see asymmetry both in and out of the disk plane.

\begin{figure}
    \centering
    \includegraphics[width=\columnwidth]{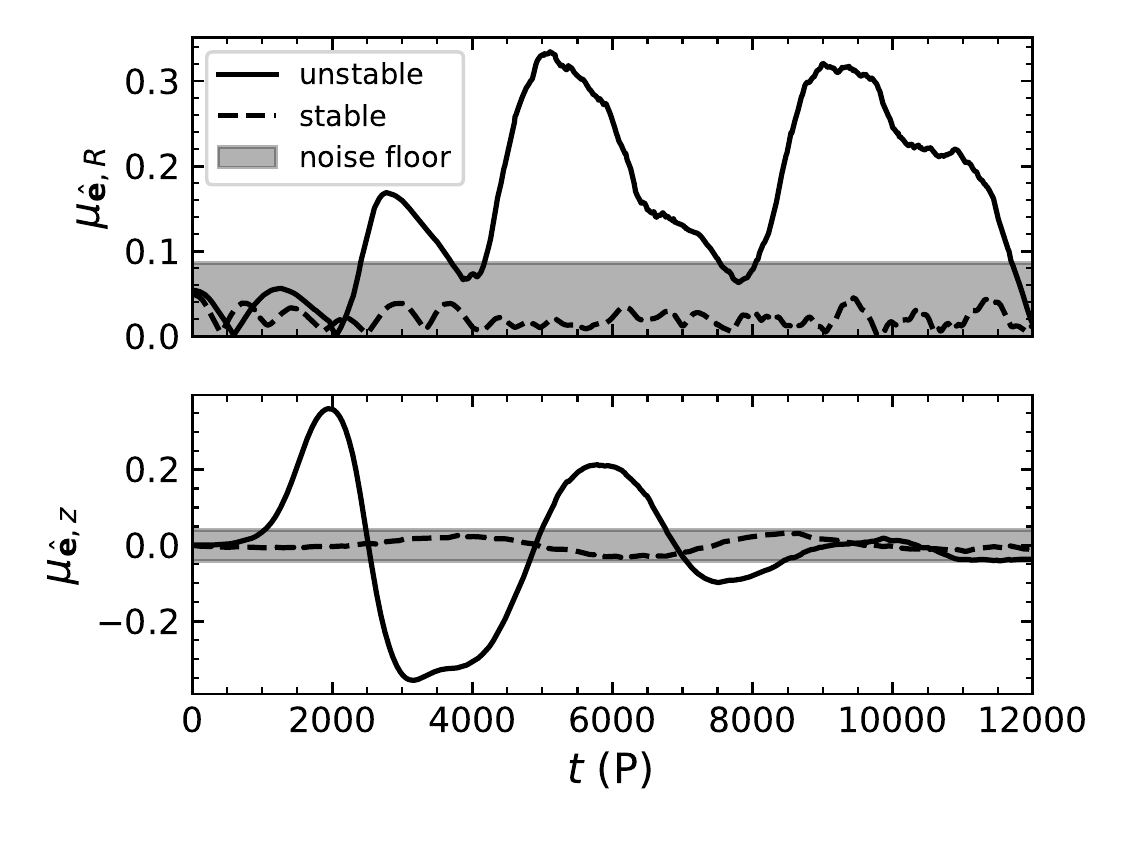}
    \caption{Length of the projection of the normalized mean eccentricity vector in the $x$-$y$ plane, $\muR$, and $z$-component of the normalized mean eccentricity vector, $\muz$, as function of time for two simulations. The grey band shows the noise floor. Both simulations have identical initial orbital distributions, but one has a strong $J_2$ moment added to the central body that suppresses the instability (stable) and the other has no added $J_2$ (unstable).}
    \label{fig:ecc-vec-evo}
\end{figure}

We quantify alignment of the orbits using the mean normed eccentricity vector, $\muvec = \sum_{i=1}^{N} \nicefrac{\hat{\bm{e}}_i}{N}$, where $\hat{\bm{e}}_i$ is a unit vector pointing from orbit $i$'s focus to pericenter.
The inclination instability reveals itself as a rapid increase in $\muz$.
The apsidal clustering that follows occurs in the $x$-$y$ plane, which we quantify using $\muR = \sqrt{\mux^2 + \muy^2}$. 

The time series evolution of $\muz$ and $\muR$ for two simulations is shown in Figure~\ref{fig:ecc-vec-evo}.
The first simulation is unstable to the inclination instability. 
This creates an \added{out-of-plane} asymmetry \added{quantified by} $\muz$ which then appears to seed an \added{in-plane} asymmetry \added{quantified by} $\muR$ as the orbits precess back through the plane. 
This \added{in-plane} over-density attracts more orbits, increasing the strength of the perturbation. 
The mode disperses, recurring some time later. 
\added{Differential precession causes the out-of-plane orbital clustering to disappear after $\sim10^{4}\,P$. We do not anticipate this clustering to reappear beyond this time as the conditions which drove the instability in the first place (low orbital inclinations, high orbital eccentricities) will no longer be met. Artificially strong two-body scattering causes the in-plane clustering to disappear after this. We expect in-plane clustering to be sustained in simulations with larger particle numbers (see Section~\ref{sec:mode-disperse} and Figure~\ref{fig:N-plot}).}

In the second simulation, the disk is made stable against the inclination instability by the addition of a zonal quadrupole ($J_2 = 3\times10^{-5}$) moment of the central body, just strong enough to suppress the inclination instability with differential apsidal precession of disk orbits \citep[see][]{Zderic2020}.
The eccentricity vector components remain below the noise floor for the entire length of the simulation.\footnote{The noise floor for $\muR$ is calculated as the 95th percentile of 100 iterations of the initial, axisymmetric $\muR$. The noise floor for $\muz$ is calculated by bootstrapping the $2\sigma$ error of $\muz$ at late times ($t \sim 10^4$~P).}
Smaller $J_2$ moments (i.e. low enough for the instability to still occur) actually increase the longevity of the apsidal clustering.

\begin{figure*}
    \centering
    \includegraphics[]{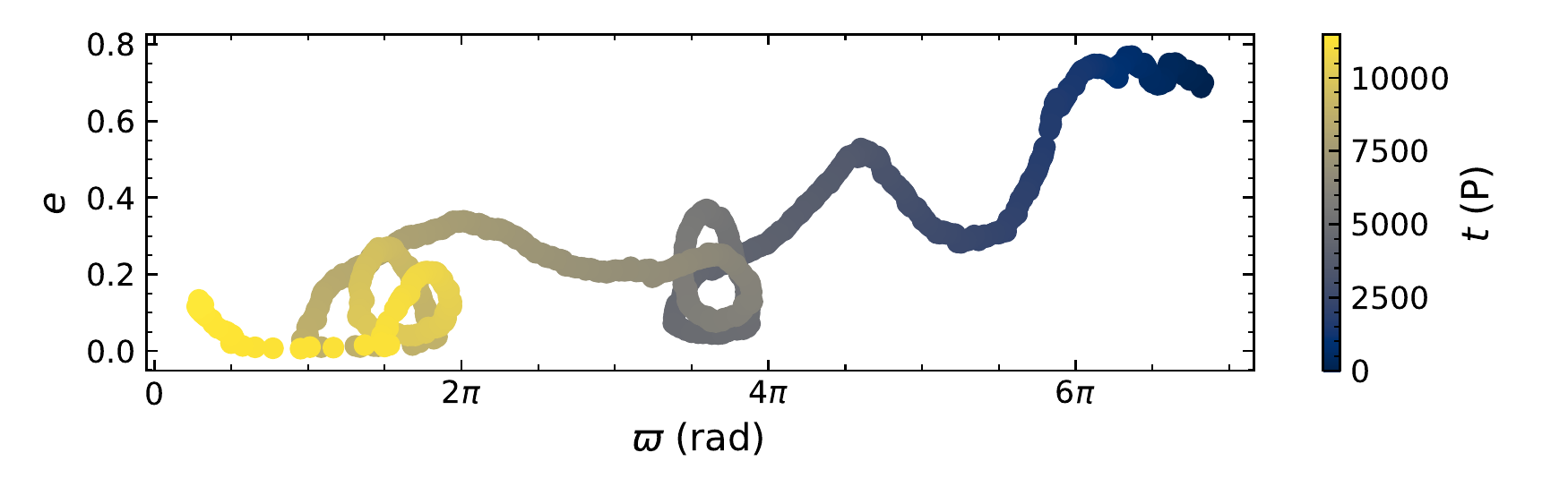}
    \caption{Evolution of the longitude of perihelion $\varpi$, and orbital eccentricity, $e$, as a function of time demonstrating the transient trapping of an orbit in an m=1 mode. 
    	The orbit precesses with retrograde motion (moving right to left) over $10^4$ orbital periods. 
	At $\sim5000~P$ and $\sim9000~P$ the orbit becomes temporarily trapped in the m=1 mode. 
	It librates within the mode, being secularly torqued by the mode to higher and lower eccentricity.}
    \label{fig:single-particle}
\end{figure*}

In Figure~\ref{fig:single-particle}, we show the time evolution of the longitude of perihelion, $\varpi$, and orbital eccentricity, $e$, of a single particle in the unstable simulation. 
The particle's orbit precesses with retrograde motion (right to left) and decreases in eccentricity during the instability. 
At $\sim5000~P$, the orbit becomes transiently trapped in the $m=1$ mode, and librates in $\varpi$-$e$ space.
After a few cycles in the mode, the particle escapes, circulates retrograde for a single cycle, and becomes transiently trapped in the mode again at $\sim9000~P$. 
Secular gravitational torques exerted on the orbit by the mode are responsible for the $\varpi$-$e$ oscillations. 
This same mechanism stabilizes eccentric nuclear disks of stars around supermassive black holes \citep{Madigan2018a}.

\section{Discussion}
\label{sec:discussion}

\subsection{Why does an $m=1$ mode develop in the plane of the disk?}

Lopsided modes in near-Keplerian disks can develop spontaneously if the disk contains a large fraction of retrograde orbits (e.g., \citealt{Touma02, Touma09, Kazandjian2013}). 
In the simulations presented here, the inclination instability produces \replaced{$\lesssim 1\%$ retrograde particles from an initially $\sim$co-planar configuration. The first orbit to reach an inclination $i > 90^{\circ}$ typically does so after apsidal alignment appears; }{retrograde orbits, but only a few ($\lesssim 1\%$) and only after apsidal alignment has already occurred.}
Thus, retrograde orbits are not responsible for the clustering observed here\added{, distinguishing these results from previous work.}

A recent series of papers \citep{Touma2019,Tremaine2020a,Tremaine2020b} show that spherical near-Keplerian potentials tend toward an ordered, lopsided state when the system is cooled below a critical temperature (directly related to RMS eccentricity of the orbits). The ordered lopsided state results from a phase transition, rather than dynamical instability, driven by resonant relaxation \citep{Rauch1996}.

Our simulations show the development of a spontaneous lopsided mode in a three-dimensional near-Keplerian potential. This is done without seeding asymmetry (beyond that from numerical noise), forcing mode development with gas dynamics, or superimposing retrograde orbits. 
The appearance of the mode appears to be contingent on the inclination instability altering the initial orbital configuration, requiring lower eccentricities, higher inclinations and clustering in arguments of perihelion.\footnote{Low eccentricity ($e < 0.5$) disks that don't undergo the instability don't show apsidal clustering, \added{nor do post-instability disks in which we deliberately randomize the arguments of perihelion}.}
Indeed, in the `stable' simulation of Figure~\ref{fig:ecc-vec-evo} in which the instability is suppressed, there is no apsidal clustering.
We hypothesize that the inclination instability \deleted{puffs ($i\uparrow$) and cools ($e\downarrow$) the system allowing} allows the system to undergo a phase transition to a lopsided state similar to the transition reported in \citet{Touma2019} for spherically symmetric systems.
\added{This hypothesis will be explored in future work.}

\subsection{Lifetime of the mode}
\label{sec:mode-disperse}

While the inclination instability appears in our compact simulations with as few as $N = 100$ particles, more particles are needed to resolve and stabilize the $m=1$ mode.
This is reminiscent of bar development in galactic disks. The bar instability in an $N$-body disk occurs on nearly identical timescales, initial mode density, and growth rate for similar disks of increasing $N$. However, if $N$ is too low, the bar will dissolve soon after formation \citep{Dubinski2009}. Similarly, $N$-body galactic disks are known to easily form recurrent short-lived, transient spirals \citep{JamesSellwood1978,Sellwood2012, Sellwood2020}. 
When particle number is increased in these simulations, the over-density reaches some minimum value and results in exponential growth of the mode (e.g., \citet{ToomreKalnajs91, Weinberg98, Sellwood2012}).  
In disk galaxies this presents as long-lasting $m=2$ spirals or an $m=2$ bar mode. 
\added{In our system, the low number of particles results in a coarse, under-populated mode. 
	Artificially strong two-body interactions perturb the osculating orbits of bodies that stream through the mode and weaken the secular torques that would trap them.}
Therefore, while our simulations show robust results, increasing particle number is well motivated. 

\begin{figure}[t!]
    \centering
    \includegraphics[width=\linewidth]{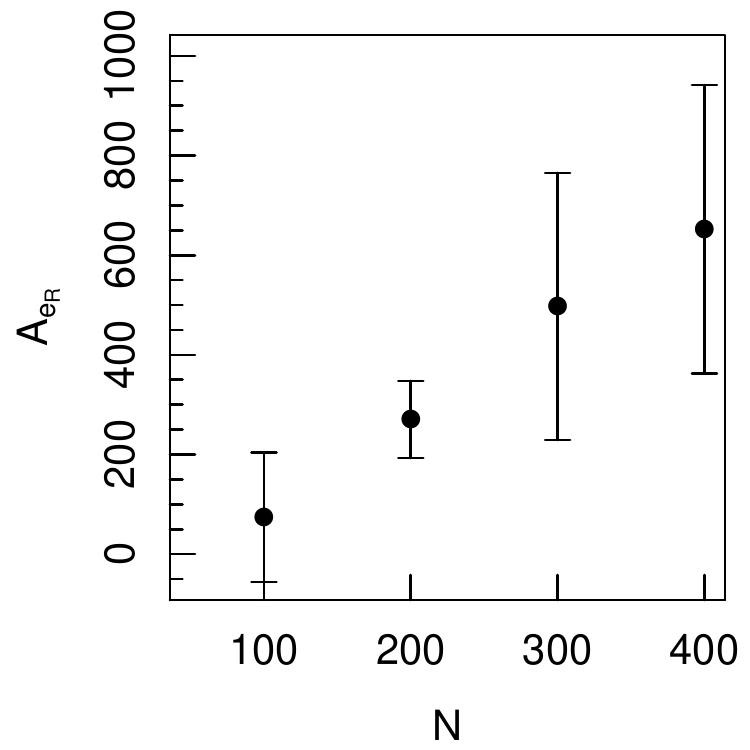}
    \caption{$A_{e_R}$,  the integral of $\muR$ above the noise floor, as a function of particle number, $N$, for a simulation length of 10,000$\,P$ .  The points and error bars show the median and standard deviation of 6 simulations in each group.}
    \label{fig:N-plot}
\end{figure}

In Figure~\ref{fig:N-plot}, we show that the strength and duration of the mode increases with $N$, the number of particles in the simulation. We do so by calculating $A_{e_R}$, the integral of $\muR$ above the noise floor over a time period of 10,000$\,P$. These initial tests indicate that the disk produces longer-lasting and stronger apsidal alignment with increasing $N$. \\
\added{At $N=400$, the mode is already quite long-lived. 
Rescaling our results for a disk mass of $\Md \sim 20\,M_\oplus$, using $a=1=100\,{\rm AU}$, $P=1000\,{\rm yr}$, the mode lasts $\sim160$ Myr.
For realistic numbers of particles, many orders of magnitude larger than what we use here, it may be reasonable to expect that the mode would be stable for the age of the solar system.}

\subsection{Application to the Solar System}

Observations reveal many unusual orbital features in the population of extreme trans-Neptunian Objects (eTNOs). 
This includes detached orbits (perihelia well beyond the orbit of Neptune), high inclinations and even retrograde orbits, clustering in arguments of perihelion $\omega$, and clustering in longitudes of perihelion $\varpi$. In particular, the clustering in $\varpi$ has been an important motivator of the Planet 9 hypothesis \citep{Batygin2016,Batygin2019}. 
For recent reviews, including in-depth discussions of observational biases, see \citet{Trujillo2020,Kavelaars2020}. 

\citet{Sefilian2019} show that test particles interacting with the potential of a thick, apsidally-aligned eccentric disk  in the outer Solar System will cluster in $\varpi$ and reproduce other key orbital features of the eTNO population. 
However, they don't directly simulate this eccentric disk or its formation\footnote{On this issue, \citet{Sefilian2019} point to a paper by Kazandjian et al. in prep.}.
Here we have shown that the late-time evolution of the inclination instability can produce such a structure from an axisymmetric disk, and that this structure is likely stable at large $N$.
The drop in eccentricity during the instability isolates the system, both from the influence of giant planets at the inner edge and external perturbations (galactic tides, passing stars) at the outer edge.  
The isolated nature of the system, and the increasing stability of the mode with $N$, suggests that the structure should be stable over a long timescale. 
The observations of Sednoids  \citep[$a \gtrsim 150 AU$, $p \gtrsim 50 AU$;][]{Brown2004,Trujillo2014} support this picture.

\section{Conclusion}
\label{sec:conclusion}

In a recent series of papers we have shown that the inclination instability in high-eccentricity, near-Keplerian disks results in high orbital inclinations, raised perihelia and $\omega$-clustering.  Here we show that the system's long-term behavior results in $\varpi$-clustering. The strength and duration of the apsidal clustering increases with increasing $N$. 
We find that both $\omega$-clustering and $\varpi$-clustering can occur at the same time. 
In the context of the solar system, the collective gravity of eTNOs could explain the observed $\omega$-clustering, $\varpi$-clustering, detached objects, and even a perihelion gap \citep[see][]{Zderic2020}.

In this letter, we present results from simulations with highly idealized initial conditions for both the simplicity of analysis and tractable computational expense. In future, we plan to simulate the long-term evolution of a high-mass primordial scattered disk including the presence of the giant planets at high-$N$. We are working on the modification of existing codes which will allow us to advance in this direction. \\

\acknowledgments
\section*{Acknowledgements}

We thank Aleksey Generozov for suggesting we use the eccentricity vector to quantify apsidal clustering. 
AM gratefully acknowledges support from the David and Lucile Packard Foundation.
This work was supported by a NASA Solar System Workings grant (80NSSC17K0720) and a NASA Earth and Space Science Fellowship (80NSSC18K1264). 
This work utilized resources from the University of Colorado Boulder Research Computing Group, which is supported by the National Science Foundation (awards ACI-1532235 and ACI-1532236), the University of Colorado Boulder, and Colorado State University. 

\software{\texttt{REBOUND} \citep{Rein2012}, 
\texttt{REBOUNDX} \citep{Tamayo2019}}

\footnotesize{
\bibliographystyle{aastex}
\bibliography{master}
}

\end{document}